# Gravity gradient torque of spacecraft orbiting asteroids

*Yue Wang, Shijie Xu*

Department of Guidance, Navigation and Control, School of Astronautics, Beijing University of Aeronautics and Astronautics, Beijing, China

## Abstract

**Purpose** – This paper presents a full fourth-order model of the gravity gradient torque of spacecraft around asteroids by taking into consideration of the inertia integrals of the spacecraft up to the fourth order, which is an improvement of the previous fourth-order model of the gravity gradient torque.

**Design/methodology/approach** – The fourth-order gravitational potential of the spacecraft is derived based on Taylor expansion. Then the expression of the gravity gradient torque in terms of gravitational potential derivatives is derived. By using the formulation of the gravitational potential, explicit formulations of the full fourth-order gravity gradient torque are obtained. Then a numerical simulation is carried out to verify our model.

**Findings** – We find that our model is more sound and precise than the previous fourth-order model due to the consideration of higher-order inertia integrals of the spacecraft. Numerical simulation results show that the motion of the previous fourth-order model is quite different from the exact motion, while our full fourth-order model fits the exact motion very well. Our full fourth-order model is precise enough for high-precision attitude dynamics and control around asteroids.

**Practical implications** – This high-precision model is of importance for the future asteroids missions for scientific explorations and near-Earth objects mitigation.

**Originality/value** – In comparison with the previous model, a gravity gradient torque model around asteroids that is more sound and precise is established. This model is valuable for high-precision attitude dynamics and control around asteroids.

**Keywords** Spacecraft; Gravity gradient torque; Asteroids; 2nd degree and order-gravity field; Inertia integral

**Paper type** Research paper

## Introduction

Studies on asteroids could provide answers to fundamental questions concerning the past of our Solar System. Over the last two decades, the interest in spacecraft missions to asteroids has increased. The spacecraft can make near observations, bring back samples from the asteroids and provide more detailed information than ground-based observations. Several missions have been developed with big successes, such as NASA's Near Earth Asteroid Rendezvous (NEAR) mission to the asteroid Eros and the JAXA (Japanese) mission Hayabusa to the asteroid Itokawa. Several other missions are currently under development.

One of the key elements in designing such a mission is the analysis of the dynamical behavior of spacecraft around asteroids. The irregular shape, non-spherical mass distribution and rotational state of the asteroid make the dynamics of the spacecraft quite different from that around the Earth. Therefore, it is necessary to investigate the orbital and attitude dynamics around asteroids in details. The orbital dynamics around asteroids has been studied in many papers (for a recent review see Kumar, 2008), while the attitude dynamics around asteroids has been studied by Riverin and Misra (2002), Misra and Panchenko (2006) and Kumar (2008). The gravity gradient torque is the main perturbation of the attitude motion. In this paper, we focus on the gravity gradient torque of the spacecraft in the non-central gravity field of asteroids.

The gravity gradient torque of spacecraft about non-spherical bodies, such as the Earth, has been studied in several works, such as Sarychev (1961), Schlegel (1966) and Hughes (1986). Their results showed that the main term of the gravity gradient torque was contributed by the central component of the gravity field of the Earth. The oblateness of the Earth makes a contribution to the gravity gradient torque, which is approximately 5 orders of magnitude less than the main term on the geosynchronous orbit (Kumar, 2008). This is the reason why the oblateness of the Earth is not taken into consideration in the attitude dynamics of spacecraft around the Earth in theoretical studies and practical applications. However, as shown by Riverin and Misra (2002), Misra and Panchenko (2006) and Kumar (2008) the effects of the non-central gravity field of the asteroids on the attitude motion, which can be very significant, should be taken into consideration.

In these previous studies on the attitude dynamics in a non-central gravity field (Riverin and Misra, 2002, Misra and Panchenko, 2006, Kumar, 2008), inertia integrals of the spacecraft up to the second order were considered. However, the third and fourth-order inertia integrals, which appear in lower-order terms of the gravity gradient torque than the non-central component of the gravity field, were not considered. As a result, only the second-order terms and parts of the fourth-order terms were included in the gravity gradient torque, with the third-order terms and other fourth-order terms neglected. Thus, the previous model of the gravity gradient torque can be improved by taking into consideration of the higher-order inertia integrals of the spacecraft.

In this paper, by taking into consideration of the inertia integrals of the spacecraft up to the fourth order, a full fourth-order model of the gravity gradient torque around asteroids is established. The gravity field of the asteroid is considered to be a 2nd degree and order-gravity field with harmonic coefficients $C_{20}$ and $C_{22}$. According to the conclusions by Wang and Xu (2012), this approximation is precise enough for a fourth-order gravity gradient torque model. The fourth-order gravitational potential of the spacecraft is derived based on Taylor expansion. Then the expression of the gravity gradient torque in terms of gravitational potential derivatives is derived. By using the formulation of the fourth-order gravitational potential, explicit formulations of the full fourth-order gravity gradient torque are obtained. Some useful conclusions are reached. Then a numerical simulation, in which a special spacecraft consisted of 36 point masses is considered, is carried out to verify our model.

**Statement of the Problem**

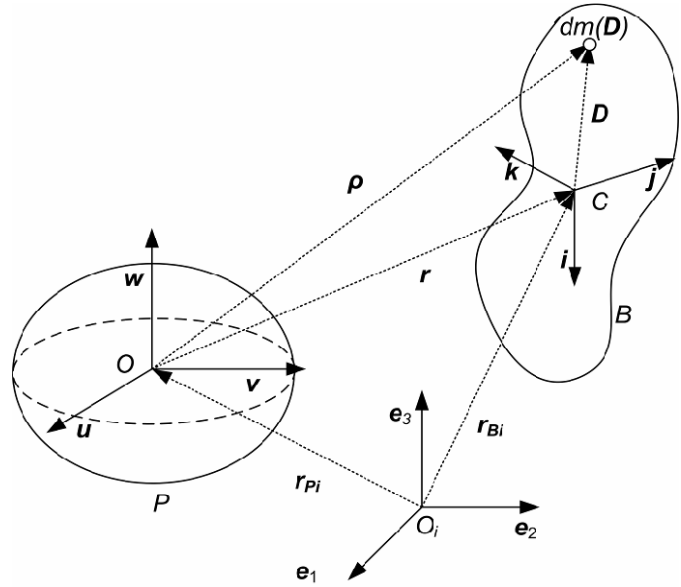

**Figure 1    A rigid spacecraft B around the asteroid P.**

As described in Figure 1, we consider a rigid spacecraft $B$ moving around the asteroid $P$. The inertial reference frame is given by $S_i=\{e_1, e_2, e_3\}$ with $O_i$ as its origin. The body-fixed reference frames of the asteroid and the spacecraft are given by $S_P=\{u, v, w\}$ and $S_B=\{i, j, k\}$ with $O$ and $C$ as their origins respectively. The origin of the frame $S_P$ is at the mass center of the asteroid, and the coordinate axes are chosen to be aligned along the principal moments of inertia of the asteroid. The principal moments of inertia of the asteroid are assumed to satisfy the following inequation

$$I_{P,zz} > I_{P,yy} > I_{P,xx}. \tag{1}$$

Then the 2nd degree and order-gravity field of the asteroid can be represented by the harmonic coefficients $C_{20}$ and $C_{22}$ with other harmonic coefficients vanished. The harmonic coefficients $C_{20}$ and $C_{22}$ are defined by

$$C_{20} = -\frac{1}{2Ma_e^2}\left(2I_{P,zz} - I_{P,xx} - I_{P,yy}\right) < 0, \tag{2}$$

$$C_{22} = \frac{1}{4Ma_e^2}\left(I_{P,yy} - I_{P,xx}\right) > 0, \tag{3}$$

where $M$ and $a_e$ are the mass and the mean equatorial radius of the asteroid respectively. Also the frame $S_B$ is attached to the mass center of the spacecraft and coincides with the principal axes reference frame of the spacecraft.

The attitude matrices of $S_P$ and $S_B$ with respect to the inertial frame $S_i$ are denoted by $A_P$ and $A_B$ respectively

$$A_P = [\boldsymbol{u}_i, \boldsymbol{v}_i, \boldsymbol{w}_i] = \begin{bmatrix} u_i^x & v_i^x & w_i^x \\ u_i^y & v_i^y & w_i^y \\ u_i^z & v_i^z & w_i^z \end{bmatrix} \in SO(3), \quad A_B = [\boldsymbol{i}_i, \boldsymbol{j}_i, \boldsymbol{k}_i] = \begin{bmatrix} i_i^x & j_i^x & k_i^x \\ i_i^y & j_i^y & k_i^y \\ i_i^z & j_i^z & k_i^z \end{bmatrix} \in SO(3), \quad (4)$$

where $\boldsymbol{u}_i$, $\boldsymbol{v}_i$, $\boldsymbol{w}_i$, $\boldsymbol{i}_i$, $\boldsymbol{j}_i$ and $\boldsymbol{k}_i$ are coordinates of the unit vectors $\boldsymbol{u}$, $\boldsymbol{v}$, $\boldsymbol{w}$, $\boldsymbol{i}$, $\boldsymbol{j}$ and $\boldsymbol{k}$ in the inertial frame $S_i$ respectively. $SO(3)$ is the 3-dimensional special orthogonal group. The matrices $A_P$ and $A_B$ are also the coordinate transformation matrices from the corresponding body-fixed frame to the inertial frame $S_i$. The relative attitude matrix of the spacecraft with respect to the asteroid is given by

$$C = A_P^T A_B. \quad (5)$$

The attitude matrices $A_P$, $A_B$ and $C$ can also be written as follows

$$A_P = [\boldsymbol{\alpha}_P, \boldsymbol{\beta}_P, \boldsymbol{\gamma}_P]^T = \begin{bmatrix} \alpha_P^x & \alpha_P^y & \alpha_P^z \\ \beta_P^x & \beta_P^y & \beta_P^z \\ \gamma_P^x & \gamma_P^y & \gamma_P^z \end{bmatrix}, \quad A_B = [\boldsymbol{\alpha}_B, \boldsymbol{\beta}_B, \boldsymbol{\gamma}_B]^T = \begin{bmatrix} \alpha_B^x & \alpha_B^y & \alpha_B^z \\ \beta_B^x & \beta_B^y & \beta_B^z \\ \gamma_B^x & \gamma_B^y & \gamma_B^z \end{bmatrix}, \quad (6)$$

$$C = [\boldsymbol{\alpha}, \boldsymbol{\beta}, \boldsymbol{\gamma}]^T = \begin{bmatrix} \alpha^x & \alpha^y & \alpha^z \\ \beta^x & \beta^y & \beta^z \\ \gamma^x & \gamma^y & \gamma^z \end{bmatrix}, \quad (7)$$

where $\boldsymbol{\alpha}$, $\boldsymbol{\beta}$ and $\boldsymbol{\gamma}$ are coordinates of the unit vectors $\boldsymbol{u}$, $\boldsymbol{v}$ and $\boldsymbol{w}$ in the body-fixed frame of the spacecraft $S_B$. The matrix $C$ is the coordinate transformation matrix from the frame $S_B$ to the body-fixed frame of the asteroid $S_P$.

$\boldsymbol{r}_{Pi}$ and $\boldsymbol{r}_{Bi}$ are radius vectors of the mass center of the asteroid $O$ and the mass center of the spacecraft $C$ with respect to $O_i$ expressed in the inertial frame $S_i$ respectively. Then the radius vector of the mass center $C$ with respect to the mass center $O$ expressed in the body-fixed frame $S_P$, denoted by $\boldsymbol{r}$, can be calculated by

$$\boldsymbol{r} = A_P^T (\boldsymbol{r}_{Bi} - \boldsymbol{r}_{Pi}). \quad (8)$$

$\boldsymbol{D}$ is the radius vector of the mass element $dm(\boldsymbol{D})$ of the spacecraft with respect to the mass center $C$ expressed in the body-fixed frame $S_B$. Then the radius vector of the mass element $dm(\boldsymbol{D})$ with respect to the mass center of the asteroid $O$ expressed in the body-fixed frame of the asteroid $S_P$, denoted by $\boldsymbol{\rho}$, is given by

$$\boldsymbol{\rho} = \boldsymbol{r} + C\boldsymbol{D}. \quad (9)$$

According to Eqn (5), the vectors $\boldsymbol{\alpha}$, $\boldsymbol{\beta}$ and $\boldsymbol{\gamma}$ can be written in terms of $A_P$ and $A_B$ as follows

$$\boldsymbol{\alpha} = \alpha_P^x \boldsymbol{\alpha}_B + \beta_P^x \boldsymbol{\beta}_B + \gamma_P^x \boldsymbol{\gamma}_B, \quad (10)$$

$$\boldsymbol{\beta} = \alpha_P^y \boldsymbol{\alpha}_B + \beta_P^y \boldsymbol{\beta}_B + \gamma_P^y \boldsymbol{\gamma}_B, \tag{11}$$

$$\boldsymbol{\gamma} = \alpha_P^z \boldsymbol{\alpha}_B + \beta_P^z \boldsymbol{\beta}_B + \gamma_P^z \boldsymbol{\gamma}_B. \tag{12}$$

**Figure 2   A unit mass point particle in the gravity field of the asteroid.**

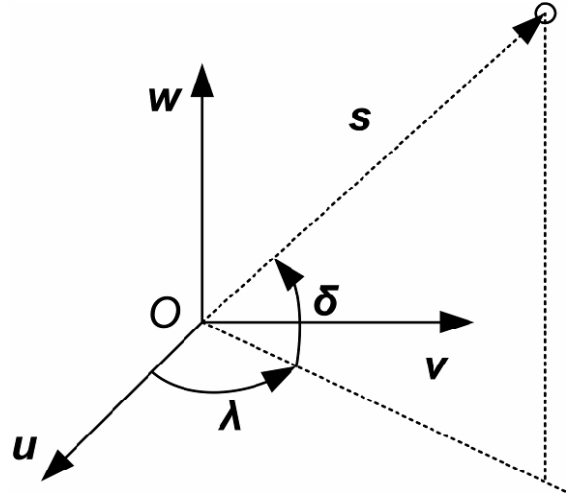

As described in Figure 2, the gravitational potential of a unit mass point particle in the gravity field of the asteroid is given by

$$V_{UMP} = -\frac{\mu}{s} - \frac{\mu}{s^3}\left[\tau_0\left(1 - \frac{3}{2}\cos^2\delta\right) + 3\tau_2\cos^2\delta\cos 2\lambda\right], \tag{13}$$

where $\mu = GM$, $G$ is the Gravitational Constant, $\tau_0 = a_e^2 C_{20}$, $\tau_2 = a_e^2 C_{22}$, $s$ is the distance of the particle from the mass center of the asteroid, $\lambda$ and $\delta$ are longitude and latitude of the particle respectively. The longitude $\lambda$ is measured counterclockwise from the *u*-axis in the *u-v* plane, and the latitude $\delta$ is measured from the *u-v* plane towards the *w*-axis. Eqn (13) can be also written in terms of *x*, *y* and *z*, the components of the position vector of the particle $s$, as follows

$$V_{UMP} = -\frac{\mu}{s}\left[1 + \frac{\tau_0}{s^2}\left(\frac{3}{2}\frac{z^2}{s^2} - \frac{1}{2}\right) + \frac{3\tau_2}{s^2}\frac{x^2 - y^2}{s^2}\right]. \tag{14}$$

The gravitational force of the unit mass point particle can be calculated by first order partial derivatives of the gravitational potential $V_{UMP}$.

## Mutual Gravitational Potential

According to Eqns (9) and (14), the gravitational potential of the mass element $dm(\boldsymbol{D})$ is given by

$$dV = -\frac{\mu dm(\boldsymbol{D})}{\rho}\left[1 + \frac{\tau_0}{\rho^2}\left(\frac{3}{2}\frac{(\rho^z)^2}{\rho^2} - \frac{1}{2}\right) + 3\tau_2\frac{(\rho^x)^2 - (\rho^y)^2}{\rho^4}\right]. \tag{15}$$

Using Eqn (9) and the coordinate transformation matrix $C^T$, we have

$$P = C^T \rho = C^T r + D = R + D, \qquad (16)$$

where $P$ and $R$ are components of the vectors $\rho$ and $r$ in the body-fixed frame of the spacecraft $S_B$ respectively. Then $\rho$, $\rho^x$, $\rho^y$ and $\rho^z$ can be written in terms of $R + D$ and the matrix $C$ as follows

$$\rho = |R+D|, \quad \rho^x = \alpha \cdot (R+D), \quad \rho^y = \beta \cdot (R+D), \quad \rho^z = \gamma \cdot (R+D). \qquad (17)$$

Therefore, the gravitational potential of the mass element $dm(D)$ can be written as

$$dV = -\frac{\mu dm(D)}{|R+D|}\left[1 + \frac{\tau_0}{|R+D|^2}\left(\frac{3(\gamma \cdot (R+D))^2}{2|R+D|^2} - \frac{1}{2}\right) + 3\tau_2 \frac{(\alpha \cdot (R+D))^2 - (\beta \cdot (R+D))^2}{|R+D|^4}\right]. \qquad (18)$$

The gravitational potential of the spacecraft is formulated by the integration over $B$

$$V(R, \alpha, \beta, \gamma) = \int_B dV(R, \alpha, \beta, \gamma, D). \qquad (19)$$

According to $|R+D|^2 = R^2 + D^2 + 2R \cdot D$, $|R+D|$ can be written as

$$|R+D| = R\left(1 + \frac{2\bar{R} \cdot D}{R} + \frac{D^2}{R^2}\right)^{\frac{1}{2}}, \qquad (20)$$

where $\bar{R}$ is the unit vector along the vector $R$. By using Eqn (20), the expressions about $|R+D|$ in Eqn (18) can be written in the form of series through Taylor expansion and truncation on the appropriate order

$$\frac{1}{|R+D|} = \frac{1}{R} - \frac{\bar{R} \cdot D}{R^2} + \left[\frac{3}{2}\frac{(\bar{R} \cdot D)^2}{R^3} - \frac{1}{2}\frac{D^2}{R^3}\right] + \left[\frac{3}{2}\frac{D^2(\bar{R} \cdot D)}{R^4} - \frac{5}{2}\frac{(\bar{R} \cdot D)^3}{R^4}\right]$$
$$+ \left[\frac{3}{8}\frac{D^4}{R^5} - \frac{15}{4}\frac{D^2(\bar{R} \cdot D)^2}{R^5} + \frac{35}{8}\frac{(\bar{R} \cdot D)^4}{R^5}\right] + O(R^{-6}), \qquad (21)$$

$$\frac{1}{|R+D|^3} = \frac{1}{R^3} - 3\frac{\bar{R} \cdot D}{R^4} + \left[\frac{15}{2}\frac{(\bar{R} \cdot D)^2}{R^5} - \frac{3}{2}\frac{D^2}{R^5}\right] + O(R^{-6}), \qquad (22)$$

$$\frac{(\alpha \cdot (R+D))^2}{|R+D|^5} = \frac{(\alpha \cdot \bar{R})^2}{R^3} + \left(\frac{2(\alpha \cdot \bar{R})(\alpha \cdot D)}{R^4} - 5\frac{(\alpha \cdot \bar{R})^2(\bar{R} \cdot D)}{R^4}\right)$$
$$+ \left[\frac{(\alpha \cdot D)^2}{R^5} - \frac{5}{2}\frac{(\alpha \cdot \bar{R})^2 D^2}{R^5} + \frac{35}{2}\frac{(\alpha \cdot \bar{R})^2(\bar{R} \cdot D)^2}{R^5} - \frac{10(\alpha \cdot \bar{R})(\alpha \cdot D)(\bar{R} \cdot D)}{R^5}\right] + O(R^{-6}), \qquad (23)$$

$$\frac{(\boldsymbol{\beta}\cdot(\boldsymbol{R}+\boldsymbol{D}))^2}{|\boldsymbol{R}+\boldsymbol{D}|^5} = \frac{(\boldsymbol{\beta}\cdot\bar{\boldsymbol{R}})^2}{R^3} + \left(\frac{2(\boldsymbol{\beta}\cdot\bar{\boldsymbol{R}})(\boldsymbol{\beta}\cdot\boldsymbol{D})}{R^4} - 5\frac{(\boldsymbol{\beta}\cdot\bar{\boldsymbol{R}})^2(\bar{\boldsymbol{R}}\cdot\boldsymbol{D})}{R^4}\right)$$
$$+ \left[\frac{(\boldsymbol{\beta}\cdot\boldsymbol{D})^2}{R^5} - \frac{5}{2}\frac{(\boldsymbol{\beta}\cdot\bar{\boldsymbol{R}})^2 D^2}{R^5} + \frac{35}{2}\frac{(\boldsymbol{\beta}\cdot\bar{\boldsymbol{R}})^2(\bar{\boldsymbol{R}}\cdot\boldsymbol{D})^2}{R^5} - \frac{10(\boldsymbol{\beta}\cdot\bar{\boldsymbol{R}})(\boldsymbol{\beta}\cdot\boldsymbol{D})(\bar{\boldsymbol{R}}\cdot\boldsymbol{D})}{R^5}\right] + O(R^{-6}), \quad (24)$$

$$\frac{(\boldsymbol{\gamma}\cdot(\boldsymbol{R}+\boldsymbol{D}))^2}{|\boldsymbol{R}+\boldsymbol{D}|^5} = \frac{(\boldsymbol{\gamma}\cdot\bar{\boldsymbol{R}})^2}{R^3} + \left(\frac{2(\boldsymbol{\gamma}\cdot\bar{\boldsymbol{R}})(\boldsymbol{\gamma}\cdot\boldsymbol{D})}{R^4} - 5\frac{(\boldsymbol{\gamma}\cdot\bar{\boldsymbol{R}})^2(\bar{\boldsymbol{R}}\cdot\boldsymbol{D})}{R^4}\right)$$
$$+ \left[\frac{(\boldsymbol{\gamma}\cdot\boldsymbol{D})^2}{R^5} - \frac{5}{2}\frac{(\boldsymbol{\gamma}\cdot\bar{\boldsymbol{R}})^2 D^2}{R^5} + \frac{35}{2}\frac{(\boldsymbol{\gamma}\cdot\bar{\boldsymbol{R}})^2(\bar{\boldsymbol{R}}\cdot\boldsymbol{D})^2}{R^5} - \frac{10(\boldsymbol{\gamma}\cdot\bar{\boldsymbol{R}})(\boldsymbol{\gamma}\cdot\boldsymbol{D})(\bar{\boldsymbol{R}}\cdot\boldsymbol{D})}{R^5}\right] + O(R^{-6}). \quad (25)$$

By using Eqns (21)-(25), the leading terms of $dV$ up to the fourth order can be written as

$$dV^{(0)} = -\frac{\mu dm(\boldsymbol{D})}{R}, \quad (26)$$

$$dV^{(1)} = \mu\frac{\bar{\boldsymbol{R}}\cdot\boldsymbol{D}}{R^2}dm(\boldsymbol{D}), \quad (27)$$

$$dV^{(2)} = -\frac{\mu dm(\boldsymbol{D})}{2R^3}\left[3(\bar{\boldsymbol{R}}\cdot\boldsymbol{D})^2 - D^2 - \tau_0 + 3\tau_0(\boldsymbol{\gamma}\cdot\bar{\boldsymbol{R}})^2 + 6\tau_2(\boldsymbol{\alpha}\cdot\bar{\boldsymbol{R}})^2 - 6\tau_2(\boldsymbol{\beta}\cdot\bar{\boldsymbol{R}})^2\right], \quad (28)$$

$$dV^{(3)} = -\frac{\mu dm(\boldsymbol{D})}{2R^4}\Big\{(\bar{\boldsymbol{R}}\cdot\boldsymbol{D})\left[3D^2 + 3\tau_0 - 5(\bar{\boldsymbol{R}}\cdot\boldsymbol{D})^2\right] + 6\tau_0(\boldsymbol{\gamma}\cdot\bar{\boldsymbol{R}})(\boldsymbol{\gamma}\cdot\boldsymbol{D}) - 15\tau_0(\boldsymbol{\gamma}\cdot\bar{\boldsymbol{R}})^2(\bar{\boldsymbol{R}}\cdot\boldsymbol{D})$$
$$+ 12\tau_2(\boldsymbol{\alpha}\cdot\bar{\boldsymbol{R}})(\boldsymbol{\alpha}\cdot\boldsymbol{D}) - 30\tau_2(\boldsymbol{\alpha}\cdot\bar{\boldsymbol{R}})^2(\bar{\boldsymbol{R}}\cdot\boldsymbol{D}) - 12\tau_2(\boldsymbol{\beta}\cdot\bar{\boldsymbol{R}})(\boldsymbol{\beta}\cdot\boldsymbol{D}) + 30\tau_2(\boldsymbol{\beta}\cdot\bar{\boldsymbol{R}})^2(\bar{\boldsymbol{R}}\cdot\boldsymbol{D})\Big\}, \quad (29)$$

$$dV^{(4)} = -\frac{\mu dm(\boldsymbol{D})}{8R^5}\Big\{\left[3D^4 - 30D^2(\bar{\boldsymbol{R}}\cdot\boldsymbol{D})^2 + 35(\bar{\boldsymbol{R}}\cdot\boldsymbol{D})^4\right] + 2\tau_0\left[3D^2 - 15(\bar{\boldsymbol{R}}\cdot\boldsymbol{D})^2\right]$$
$$+ 12\tau_0(\boldsymbol{\gamma}\cdot\boldsymbol{D})^2 - 30\tau_0(\boldsymbol{\gamma}\cdot\bar{\boldsymbol{R}})^2 D^2 + 210\tau_0(\boldsymbol{\gamma}\cdot\bar{\boldsymbol{R}})^2(\bar{\boldsymbol{R}}\cdot\boldsymbol{D})^2 - 120\tau_0(\boldsymbol{\gamma}\cdot\bar{\boldsymbol{R}})(\boldsymbol{\gamma}\cdot\boldsymbol{D})(\bar{\boldsymbol{R}}\cdot\boldsymbol{D})$$
$$+ 24\tau_2(\boldsymbol{\alpha}\cdot\boldsymbol{D})^2 - 60\tau_2(\boldsymbol{\alpha}\cdot\bar{\boldsymbol{R}})^2 D^2 + 420\tau_2(\boldsymbol{\alpha}\cdot\bar{\boldsymbol{R}})^2(\bar{\boldsymbol{R}}\cdot\boldsymbol{D})^2 - 240\tau_2(\boldsymbol{\alpha}\cdot\bar{\boldsymbol{R}})(\boldsymbol{\alpha}\cdot\boldsymbol{D})(\bar{\boldsymbol{R}}\cdot\boldsymbol{D})$$
$$- 24\tau_2(\boldsymbol{\beta}\cdot\boldsymbol{D})^2 + 60\tau_2(\boldsymbol{\beta}\cdot\bar{\boldsymbol{R}})^2 D^2 - 420\tau_2(\boldsymbol{\beta}\cdot\bar{\boldsymbol{R}})^2(\bar{\boldsymbol{R}}\cdot\boldsymbol{D})^2 + 240\tau_2(\boldsymbol{\beta}\cdot\bar{\boldsymbol{R}})(\boldsymbol{\beta}\cdot\boldsymbol{D})(\bar{\boldsymbol{R}}\cdot\boldsymbol{D})\Big\}. \quad (30)$$

Substitution of Eqns (26)-(30) into Eqn (19) gives the leading terms of the gravitational potential $V$, namely $V^{(0)}$ to $V^{(4)}$. The zeroth-order gravitational potential $V^{(0)}$ is given by

$$V^{(0)} = \int_B -\frac{\mu dm(\boldsymbol{D})}{R} = -\frac{\mu m}{R}, \quad (31)$$

where $m$ is the mass of the spacecraft. Since the origin of the frame $S_B$ coincides with the mass center of the spacecraft, the first-order gravitational potential $V^{(1)}$ is vanished

$$V^{(1)} = \mu\frac{\bar{\boldsymbol{R}}}{R^2}\cdot\int_B \boldsymbol{D}dm(\boldsymbol{D}) = 0. \quad (32)$$

The inertia integrals of the spacecraft are defined by

$$J_{\underbrace{x...x}_{p-times}\underbrace{y...y}_{q-times}\underbrace{z...z}_{r-times}} = \int_B (D^x)^p (D^y)^q (D^z)^r dm(\boldsymbol{D}). \tag{33}$$

The moments of inertia are defined by $I_{xx} = J_{yy} + J_{zz}$, $I_{yy} = J_{xx} + J_{zz}$ and $I_{zz} = J_{xx} + J_{yy}$. The frame $S_B$ is the principal axes reference frame of the spacecraft, thus the product moments of inertia are all eliminated. We express $\bar{\boldsymbol{R}} \cdot \boldsymbol{D}$, $D^2$, $\boldsymbol{\alpha} \cdot \boldsymbol{D}$, $\boldsymbol{\beta} \cdot \boldsymbol{D}$ and $\boldsymbol{\gamma} \cdot \boldsymbol{D}$ in terms of the components in the body-fixed frame $S_B$ as follows

$$\bar{\boldsymbol{R}} \cdot \boldsymbol{D} = \bar{R}^x D^x + \bar{R}^y D^y + \bar{R}^z D^z, \quad D^2 = (D^x)^2 + (D^y)^2 + (D^z)^2, \tag{34}$$

$$\boldsymbol{\alpha} \cdot \boldsymbol{D} = \alpha^x D^x + \alpha^y D^y + \alpha^z D^z, \quad \boldsymbol{\beta} \cdot \boldsymbol{D} = \beta^x D^x + \beta^y D^y + \beta^z D^z, \tag{35}$$

$$\boldsymbol{\gamma} \cdot \boldsymbol{D} = \gamma^x D^x + \gamma^y D^y + \gamma^z D^z. \tag{36}$$

Using inertia integrals defined above, we can get the second-order gravitational potential $V^{(2)}$ and the third-order gravitational potential $V^{(3)}$ as follows

$$V^{(2)}(\boldsymbol{R},\boldsymbol{\alpha},\boldsymbol{\beta},\boldsymbol{\gamma}) = -\frac{\mu}{2R^3}\left[tr(\boldsymbol{I}) - 3\bar{\boldsymbol{R}}^T \boldsymbol{I}\bar{\boldsymbol{R}} - m\tau_0 + 3m\tau_0(\boldsymbol{\gamma}\cdot\bar{\boldsymbol{R}})^2 + 6m\tau_2\left((\boldsymbol{\alpha}\cdot\bar{\boldsymbol{R}})^2 - (\boldsymbol{\beta}\cdot\bar{\boldsymbol{R}})^2\right)\right], \tag{37}$$

$$V^{(3)}(\boldsymbol{R},\boldsymbol{\alpha},\boldsymbol{\beta},\boldsymbol{\gamma}) = \frac{\mu}{2R^4}\Big[\bar{R}^x\left(5(\bar{R}^x)^2 - 3\right)J_{xxx} + \bar{R}^y\left(5(\bar{R}^y)^2 - 3\right)J_{yyy} + \bar{R}^z\left(5(\bar{R}^z)^2 - 3\right)J_{zzz} + 3\bar{R}^y\left(5(\bar{R}^x)^2 - 1\right)J_{xxy}$$
$$+ 3\bar{R}^x\left(5(\bar{R}^y)^2 - 1\right)J_{xyy} + 3\bar{R}^z\left(5(\bar{R}^x)^2 - 1\right)J_{xxz} + 3\bar{R}^x\left(5(\bar{R}^z)^2 - 1\right)J_{xzz} + 3\bar{R}^z\left(5(\bar{R}^y)^2 - 1\right)J_{yyz}$$
$$+ 3\bar{R}^y\left(5(\bar{R}^z)^2 - 1\right)J_{yzz} + 30\bar{R}^x\bar{R}^y\bar{R}^z J_{xyz}\Big]. \tag{38}$$

The fourth-order gravitational potential $V^{(4)}$ can be written into three parts

$$V^{(4),C}(\boldsymbol{R},\boldsymbol{\alpha},\boldsymbol{\beta},\boldsymbol{\gamma}) = -\frac{\mu}{8R^5}\Big[\left(35(\bar{R}^x)^4 - 30(\bar{R}^x)^2 + 3\right)J_{xxxx} + \left(35(\bar{R}^y)^4 - 30(\bar{R}^y)^2 + 3\right)J_{yyyy} + 20\bar{R}^x\bar{R}^z\left(7(\bar{R}^z)^2 - 3\right)J_{xzzz}$$
$$+ 20\bar{R}^y\bar{R}^z\left(7(\bar{R}^y)^2 - 3\right)J_{yyyz} + 20\bar{R}^y\bar{R}^z\left(7(\bar{R}^z)^2 - 3\right)J_{yzzz} + \left(35(\bar{R}^z)^4 - 30(\bar{R}^z)^2 + 3\right)J_{zzzz} + 20\bar{R}^x\bar{R}^y\left(7(\bar{R}^y)^2 - 3\right)J_{xyyy}$$
$$+ 20\bar{R}^x\bar{R}^z\left(7(\bar{R}^x)^2 - 3\right)J_{xxxz} + 6\left(35(\bar{R}^x)^2(\bar{R}^y)^2 - 5(\bar{R}^x)^2 - 5(\bar{R}^y)^2 + 1\right)J_{xxyy} + 20\bar{R}^x\bar{R}^y\left(7(\bar{R}^x)^2 - 3\right)J_{xxxy}$$
$$+ 6\left(35(\bar{R}^x)^2(\bar{R}^z)^2 - 5(\bar{R}^x)^2 - 5(\bar{R}^z)^2 + 1\right)J_{xxzz} + 6\left(35(\bar{R}^y)^2(\bar{R}^z)^2 - 5(\bar{R}^y)^2 - 5(\bar{R}^z)^2 + 1\right)J_{yyzz}$$
$$+ 60\bar{R}^y\bar{R}^z\left(7(\bar{R}^x)^2 - 1\right)J_{xxyz} + 60\bar{R}^x\bar{R}^z\left(7(\bar{R}^y)^2 - 1\right)J_{xyyz} + 60\bar{R}^x\bar{R}^y\left(7(\bar{R}^z)^2 - 1\right)J_{xyzz}\Big], \tag{39}$$

$$V^{(4),C_{20}}(\boldsymbol{R},\boldsymbol{\alpha},\boldsymbol{\beta},\boldsymbol{\gamma}) = -\frac{3\mu\tau_0}{4R^5}\Big[\left(5(\boldsymbol{\gamma}\cdot\bar{\boldsymbol{R}})^2 - 1\right)tr(\boldsymbol{I}) + 5\left(1 - 7(\boldsymbol{\gamma}\cdot\bar{\boldsymbol{R}})^2\right)\bar{\boldsymbol{R}}^T \boldsymbol{I}\bar{\boldsymbol{R}} - 2\left[\boldsymbol{\gamma}^T \boldsymbol{I}\boldsymbol{\gamma} - 10(\boldsymbol{\gamma}\cdot\bar{\boldsymbol{R}})\boldsymbol{\gamma}^T \boldsymbol{I}\bar{\boldsymbol{R}}\right]\Big], \tag{40}$$

$$V^{(4),C_{22}}(\boldsymbol{R},\boldsymbol{\alpha},\boldsymbol{\beta},\boldsymbol{\gamma}) = -\frac{3\mu\tau_2}{2R^5}\Big[5\left((\boldsymbol{\alpha}\cdot\bar{\boldsymbol{R}})^2 - (\boldsymbol{\beta}\cdot\bar{\boldsymbol{R}})^2\right)\left(tr(\boldsymbol{I}) - 7\bar{\boldsymbol{R}}^T \boldsymbol{I}\bar{\boldsymbol{R}}\right)$$
$$-2\left(\boldsymbol{\alpha}^T \boldsymbol{I}\boldsymbol{\alpha} - \boldsymbol{\beta}^T \boldsymbol{I}\boldsymbol{\beta} - 10(\boldsymbol{\alpha}\cdot\bar{\boldsymbol{R}})\boldsymbol{\alpha}^T \boldsymbol{I}\bar{\boldsymbol{R}} + 10(\boldsymbol{\beta}\cdot\bar{\boldsymbol{R}})\boldsymbol{\beta}^T \boldsymbol{I}\bar{\boldsymbol{R}}\right)\Big], \tag{41}$$

$V^{(4),C}$ is the gravitational potential due to the interaction between the central component of the gravity field of the asteroid and the fourth-order inertia integrals of the spacecraft; $V^{(4),C_{20}}$ is the gravitational potential of the interaction between the second degree and zeroth order component of the gravity field and the second-order inertia integrals; $V^{(4),C_{22}}$ is due to the interaction between the second degree and second order component of the gravity field and the second-order inertia integrals.

The fourth-order approximate gravitational potential $\tilde{V}$ is the sum of $V^{(0)}$, $V^{(2)}$, $V^{(3)}$, $V^{(4),C}$, $V^{(4),C_{20}}$ and $V^{(4),C_{22}}$

$$\tilde{V}(R, \alpha, \beta, \gamma) = V^{(0)} + V^{(2)} + V^{(3)} + V^{(4),C} + V^{(4),C_{20}} + V^{(4),C_{22}}. \tag{42}$$

## Gravity Gradient Torque

After the formulation of the mutual gravitational potential between the asteroid and the spacecraft obtained, the explicit formulations of the gravity gradient torque acting on the spacecraft can be derived through the gravitational potential derivatives. The gravitational potential $\tilde{V}$ is a function of the inertial positions and attitudes of the asteroid and the spacecraft $r_{Pi}$, $r_{Bi}$, $A_P$ and $A_B$. However, according to Eqns (31) and (37)-(42), the gravitational potential $\tilde{V}$ can be determined by the relative position and attitude of the spacecraft with respect to the asteroid,

$$\tilde{V} = \tilde{V}(r_{Pi}, r_{Bi}, A_P, A_B) = \tilde{V}(R, C). \tag{43}$$

We have several relations between these two sets of variables of position and attitude, as shown by Eqns (10)-(12) and the following equation

$$R = A_B^T (r_{Bi} - r_{Pi}). \tag{44}$$

According to Maciejewski (1995), if the mutual gravitational potential $\tilde{V}$ is considered as a function of $r_{Pi}$, $r_{Bi}$, $A_P$ and $A_B$, the gravity gradient torque acting on the spacecraft $B$ expressed in the body-fixed frame $S_B$, denoted by $\tilde{T}_B$, can be calculated by

$$\tilde{T}_B = \alpha_B \times \frac{\partial \tilde{V}(r_{Pi}, r_{Bi}, A_P, A_B)}{\partial \alpha_B} + \beta_B \times \frac{\partial \tilde{V}(r_{Pi}, r_{Bi}, A_P, A_B)}{\partial \beta_B} + \gamma_B \times \frac{\partial \tilde{V}(r_{Pi}, r_{Bi}, A_P, A_B)}{\partial \gamma_B}. \tag{45}$$

Using Eqns (10)-(12) and (44), we obtain following equations through the chain rule

$$\frac{\partial \tilde{V}(r_{Pi}, r_{Bi}, A_P, A_B)}{\partial \alpha_B} = \left(\frac{\partial R}{\partial \alpha_B}\right)^T \frac{\partial \tilde{V}(R,C)}{\partial R} + \left(\frac{\partial \alpha}{\partial \alpha_B}\right)^T \frac{\partial \tilde{V}(R,C)}{\partial \alpha} + \left(\frac{\partial \beta}{\partial \alpha_B}\right)^T \frac{\partial \tilde{V}(R,C)}{\partial \beta} + \left(\frac{\partial \gamma}{\partial \alpha_B}\right)^T \frac{\partial \tilde{V}(R,C)}{\partial \gamma}$$
$$= (r_{Bi} - r_{Pi})^x \frac{\partial \tilde{V}(R,C)}{\partial R} + \alpha_P^x \frac{\partial \tilde{V}(R,C)}{\partial \alpha} + \alpha_P^y \frac{\partial \tilde{V}(R,C)}{\partial \beta} + \alpha_P^z \frac{\partial \tilde{V}(R,C)}{\partial \gamma}, \tag{46}$$

$$\frac{\partial \tilde{V}(r_{Pi}, r_{Bi}, A_P, A_B)}{\partial \beta_B} = \left(\frac{\partial R}{\partial \beta_B}\right)^T \frac{\partial \tilde{V}(R,C)}{\partial R} + \left(\frac{\partial \alpha}{\partial \beta_B}\right)^T \frac{\partial \tilde{V}(R,C)}{\partial \alpha} + \left(\frac{\partial \beta}{\partial \beta_B}\right)^T \frac{\partial \tilde{V}(R,C)}{\partial \beta} + \left(\frac{\partial \gamma}{\partial \beta_B}\right)^T \frac{\partial \tilde{V}(R,C)}{\partial \gamma}$$
$$= (r_{Bi} - r_{Pi})^y \frac{\partial \tilde{V}(R,C)}{\partial R} + \beta_P^x \frac{\partial \tilde{V}(R,C)}{\partial \alpha} + \beta_P^y \frac{\partial \tilde{V}(R,C)}{\partial \beta} + \beta_P^z \frac{\partial \tilde{V}(R,C)}{\partial \gamma}, \tag{47}$$

$$\frac{\partial \tilde{V}(\mathbf{r}_{Pi}, \mathbf{r}_{Bi}, \mathbf{A}_P, \mathbf{A}_B)}{\partial \boldsymbol{\gamma}_B} = \left(\frac{\partial \mathbf{R}}{\partial \boldsymbol{\gamma}_B}\right)^T \frac{\partial \tilde{V}(\mathbf{R}, \mathbf{C})}{\partial \mathbf{R}} + \left(\frac{\partial \boldsymbol{\alpha}}{\partial \boldsymbol{\gamma}_B}\right)^T \frac{\partial \tilde{V}(\mathbf{R}, \mathbf{C})}{\partial \boldsymbol{\alpha}} + \left(\frac{\partial \boldsymbol{\beta}}{\partial \boldsymbol{\gamma}_B}\right)^T \frac{\partial \tilde{V}(\mathbf{R}, \mathbf{C})}{\partial \boldsymbol{\beta}} + \left(\frac{\partial \boldsymbol{\gamma}}{\partial \boldsymbol{\gamma}_B}\right)^T \frac{\partial \tilde{V}(\mathbf{R}, \mathbf{C})}{\partial \boldsymbol{\gamma}}$$

$$= (\mathbf{r}_{Bi} - \mathbf{r}_{Pi})^z \frac{\partial \tilde{V}(\mathbf{R}, \mathbf{C})}{\partial \mathbf{R}} + \gamma_P^x \frac{\partial \tilde{V}(\mathbf{R}, \mathbf{C})}{\partial \boldsymbol{\alpha}} + \gamma_P^y \frac{\partial \tilde{V}(\mathbf{R}, \mathbf{C})}{\partial \boldsymbol{\beta}} + \gamma_P^z \frac{\partial \tilde{V}(\mathbf{R}, \mathbf{C})}{\partial \boldsymbol{\gamma}}, \quad (48)$$

where $\partial \mathbf{a}/\partial \mathbf{b} = \begin{bmatrix} \partial a^x/\partial \mathbf{b} & \partial a^y/\partial \mathbf{b} & \partial a^z/\partial \mathbf{b} \end{bmatrix}^T$ is the Jacobi matrix. Substitution of Eqns (46)-(48) into Eqn (45) gives

$$\tilde{\mathbf{T}}_B = \mathbf{R} \times \frac{\partial \tilde{V}(\mathbf{R}, \mathbf{C})}{\partial \mathbf{R}} + \boldsymbol{\alpha} \times \frac{\partial \tilde{V}(\mathbf{R}, \mathbf{C})}{\partial \boldsymbol{\alpha}} + \boldsymbol{\beta} \times \frac{\partial \tilde{V}(\mathbf{R}, \mathbf{C})}{\partial \boldsymbol{\beta}} + \boldsymbol{\gamma} \times \frac{\partial \tilde{V}(\mathbf{R}, \mathbf{C})}{\partial \boldsymbol{\gamma}}. \quad (49)$$

The explicit formulations of $\tilde{\mathbf{T}}_B$ are obtained by using Eqn (49) with the help of *Maple*, which are given in the Appendix. We find that every term in the gravitational potential and torque contains a product of two mass distribution parameters, among which one is of the asteroid and the other is of the spacecraft. The order of the term is sum of orders of the two parameters. For the asteroid, the zeroth-order mass distribution parameter is the mass $M$; the second-order parameters are $C_{20}$ and $C_{22}$. For the spacecraft, the mass distribution parameters are inertia integrals.

The zeroth-order inertia integral of the spacecraft, i.e. the mass, has no contribution to the gravity gradient torque, and the first-order inertia integrals of the spacecraft are vanished. Then we can conclude that the harmonic coefficients of the asteroid higher than second degree have no contribution to the fourth-order gravity gradient torque model. Therefore, the assumption of a 2nd degree and order-gravity field is precise enough for a fourth-order gravity gradient torque model. The coefficients $C_{20}$ and $C_{22}$ appear in the fourth-order terms of the gravity gradient torque along with the second-order inertia integrals of the spacecraft. These conclusions are verified by Eqns (A.1)-(A.3) in the Appendix.

The third and fourth-order inertia integrals of the spacecraft appear in the third and fourth-order terms of the gravity gradient torque along with the mass of the asteroid respectively. In previous results (Riverin and Misra, 2002, Misra and Panchenko, 2006, Kumar, 2008), the third and fourth-order inertia integrals were not considered, thus only the second-order and some fourth-order terms of the gravity gradient torque were included, with the third-order terms and parts of fourth-order terms neglected. Therefore, our full fourth-order gravity gradient torque model is more sound and precise than previous fourth-order model. This conclusion is confirmed by our numerical simulation in the next section.

## Simulation Example

A numerical simulation is carried out to verify our gravity gradient torque model. We assume that the mass center of the asteroid is stationary in the inertial space, and the asteroid is in a uniform rotation around its maximum-moment principal axis, i.e. the *w*-axis. The spacecraft is assumed to be on a stationary orbit and the orbital motion is negligibly affected by the attitude motion, as described by Figure 3.

**Figure 3    The spacecraft on a stationary orbit around the asteroid.**

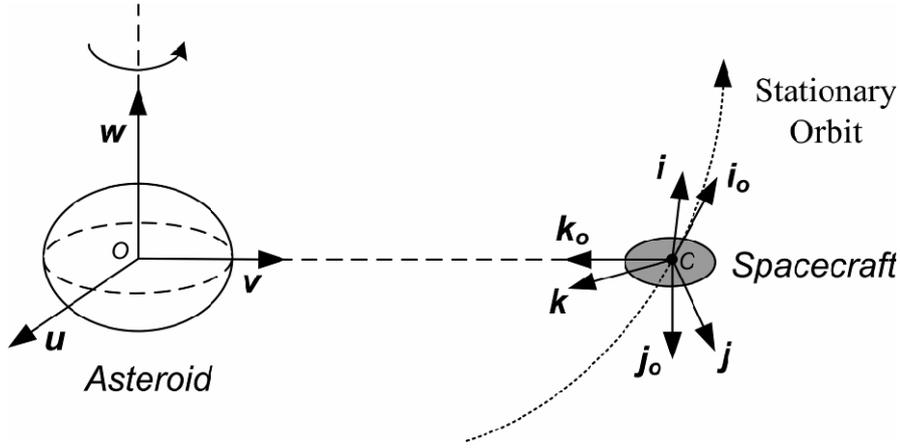

According to the orbital theory, a stationary orbit in inertial space corresponds to an equilibrium in the body-fixed frame of the asteroid. There are two kinds of stationary orbits: those that lie on the *u*-axis, and those that lie on the *v*-axis. In the numerical simulation, we consider a spacecraft located on the *v*-axis. The radius of the stationary orbit $R_S$ satisfies

$$R_S^5 - \frac{\mu}{\omega_T^2}\left(R_S^2 - \frac{3}{2}\tau_0 - 9\tau_2\right) = 0, \tag{50}$$

where $\omega_T$ is the angular velocity of the uniform rotation of the asteroid.

As described by Figure 3, the orbital reference frame is defined by $S_o=\{i_o, j_o, k_o\}$ with its origin coinciding with *C*, the mass center of the spacecraft. $k_o$ points towards the mass center of the asteroid, $j_o$ is in the opposite direction of the orbital angular momentum, and $i_o$ completes the orthogonal triad. The attitude of the spacecraft with respect to the orbital frame is defined in terms of roll, pitch and yaw angles. The sequence of rotation is: the yaw $\psi$ around the *k*-axis, followed by the pitch $\theta$ around the *j*-axis, and then the roll $\phi$ around the *i*-axis. The sequence of rotation from the frame $S_P$ to the frame $S_o$, then to the frame $S_B$ can be described as follows

$$S_P \xrightarrow{R_x(\frac{\pi}{2})} \circ \xrightarrow{R_z(\pi)} S_o \xrightarrow{R_z(\psi)} \circ \xrightarrow{R_y(\theta)} \circ \xrightarrow{R_x(\phi)} S_B.$$

The coordinate transformation matrix $L_{OP}$ from the body-fixed frame of the asteroid $S_P$ to the orbital frame $S_o$ and the coordinate transformation matrix $L_{BO}$ from the orbital frame $S_o$ to the body-fixed frame of the spacecraft $S_B$ can be calculated according to the sequence of rotation given above and Hughes (1986).

We assume further that the gravity gradient torque is the only external torque acting on the spacecraft. For the equations of the attitude motion see Hughes (1986). Notice that the relative angular velocity of the spacecraft with respect to the orbital frame $S_o$ expressed in the body-fixed frame $S_B$, $\boldsymbol{\Omega}_r = \begin{bmatrix} \Omega_r^x & \Omega_r^y & \Omega_r^z \end{bmatrix}^T$ can be calculated by

$$\boldsymbol{\Omega}_r = \boldsymbol{\Omega} - \boldsymbol{L}_{BO}\boldsymbol{\Omega}_{Orbit} = \boldsymbol{\Omega} - \boldsymbol{L}_{BO}\begin{bmatrix} 0 & -\omega_T & 0 \end{bmatrix}^T, \tag{51}$$

where $\boldsymbol{\Omega} = \begin{bmatrix} \Omega^x & \Omega^y & \Omega^z \end{bmatrix}^T$ is the angular velocity of the spacecraft expressed in the body-fixed frame $S_B$, and $\boldsymbol{\Omega}_{Orbit}$ is the angular velocity of the orbital frame $S_o$ expressed in itself.

With the explicit formulations of the gravity gradient torque given in Appendix, the system of differential equations governing the attitude motion is autonomous. Numerical simulations can be performed.

The parameters of the asteroid and its gravity field are assumed to be as follows: $M = 1.4091 \times 10^{12}\,\text{kg}$, $\tau_0 = -7.275 \times 10^4\,\text{m}^2$, $\tau_2 = 1.263 \times 10^4\,\text{m}^2$, and $\omega_T = 1.7453 \times 10^{-4}\,\text{s}^{-1}$. The radius of the stationary orbit $R_S$ is equal to 1454.952m by Eqn (50).

Here we consider a special spacecraft that is consisted of 36 point masses connected by rigid massless rods, as shown by Figure 4. The mass of each point mass is assumed to be 100kg, and the unit of length in Figure 4 is meter. With the position of every point mass in the body-fixed frame $S_B$ already known, the inertia integrals of the spacecraft can be calculated easily through Eqn (33). The gravitational force of each point mass can be calculated by the first order partial derivatives of the gravitational potential $V_{UMP}$, and then the exact gravity gradient torque can be obtained by adding the gravitational torque of each point mass with respect to the mass center $C$. Thus we can make comparisons between the motions of the previous fourth-order model, our full fourth-order model and the exact model. Through these comparisons, different approximate models can be evaluated.

**Figure 4** A special spacecraft consisted of 36 point masses connected by rigid massless rods.

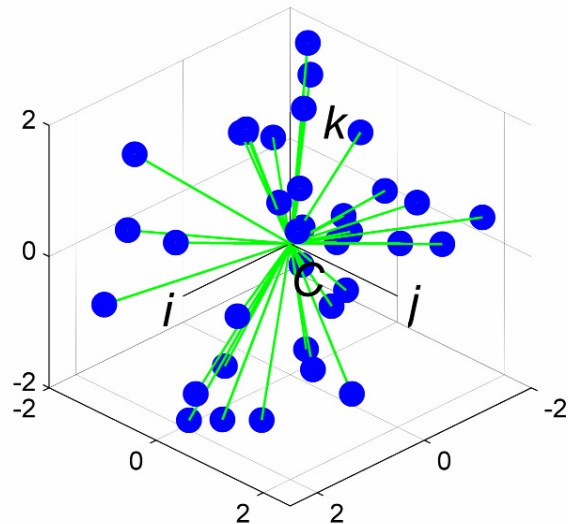

The initial conditions of the numerical simulation are set as that the Euler angles $\psi$, $\theta$ and $\phi$ are all zero, and the spacecraft has the same angular velocity as the orbital frame, i.e. $\boldsymbol{\Omega}_r = \boldsymbol{0}$. The time histories of the yaw, pitch and roll motions of the spacecraft are given in Figures 5, 6 and 7 respectively. Our full fourth-order gravity gradient torque model is denoted by *FourthOrder* in these figures, and the previous fourth-order model in Riverin and Misra (2002), Misra and Panchenko (2006) and Kumar (2008), is denoted by *PreFourthOrder*. *SecondOrder* is the second-order part of the fourth-order model, i.e. the usual gravity gradient torque model in the traditional spacecraft attitude dynamics in a central gravity field, and *Precise* is the exact model.

**Figure 5** Yaw motion with three gravity gradient torque models and exact gravity gradient torque.

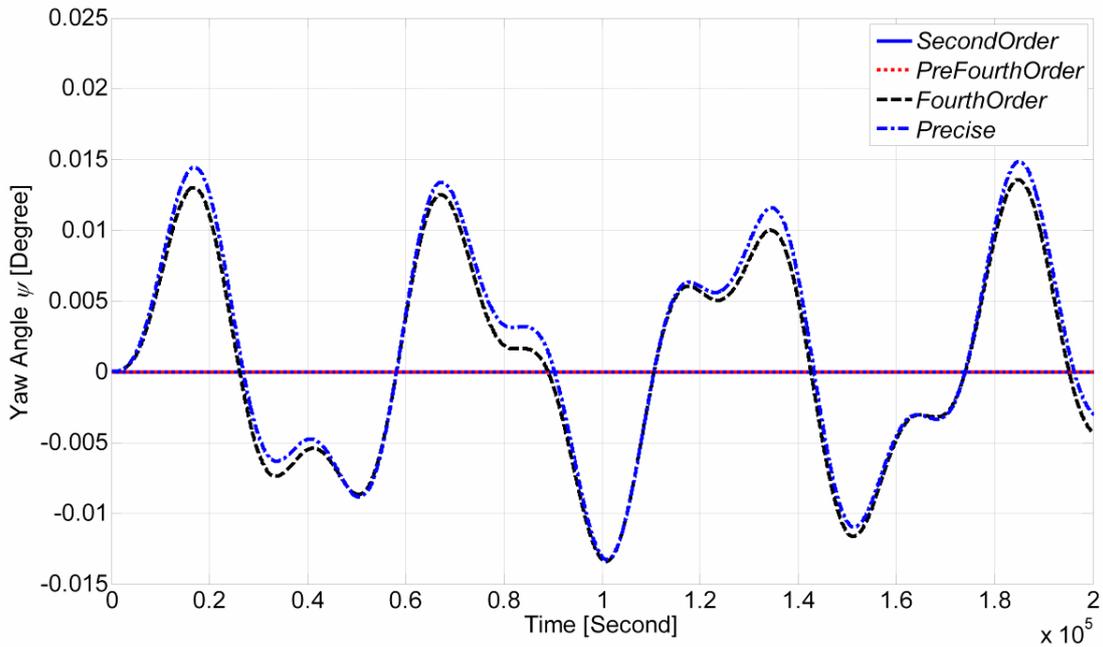

**Figure 6** Pitch motion with three gravity gradient torque models and exact gravity gradient torque.

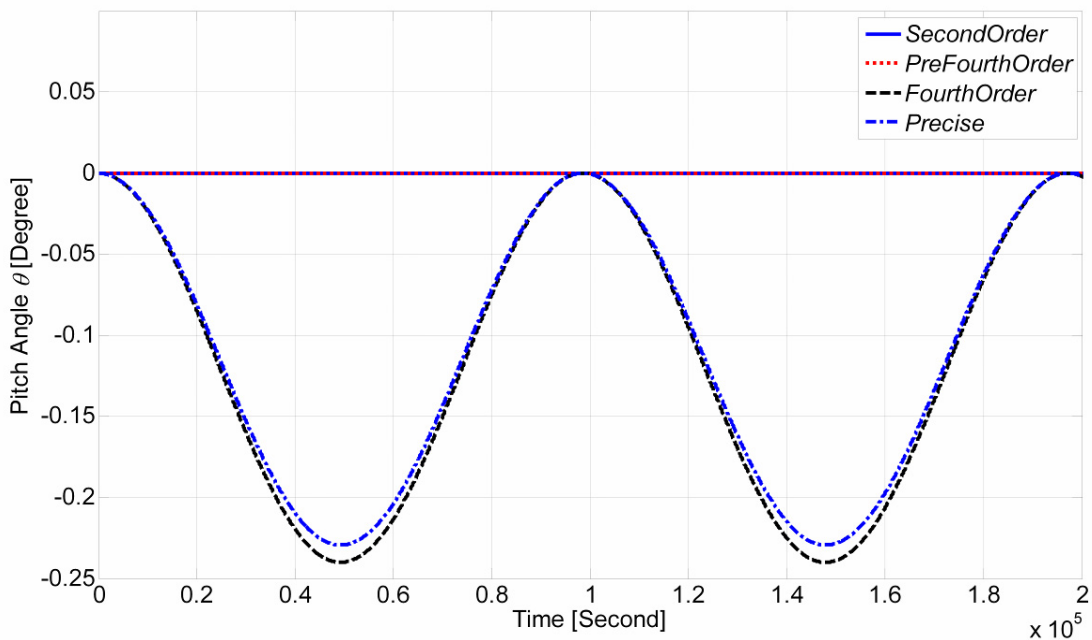

**Figure 7    Roll motion with three gravity gradient torque models and exact gravity gradient torque.**

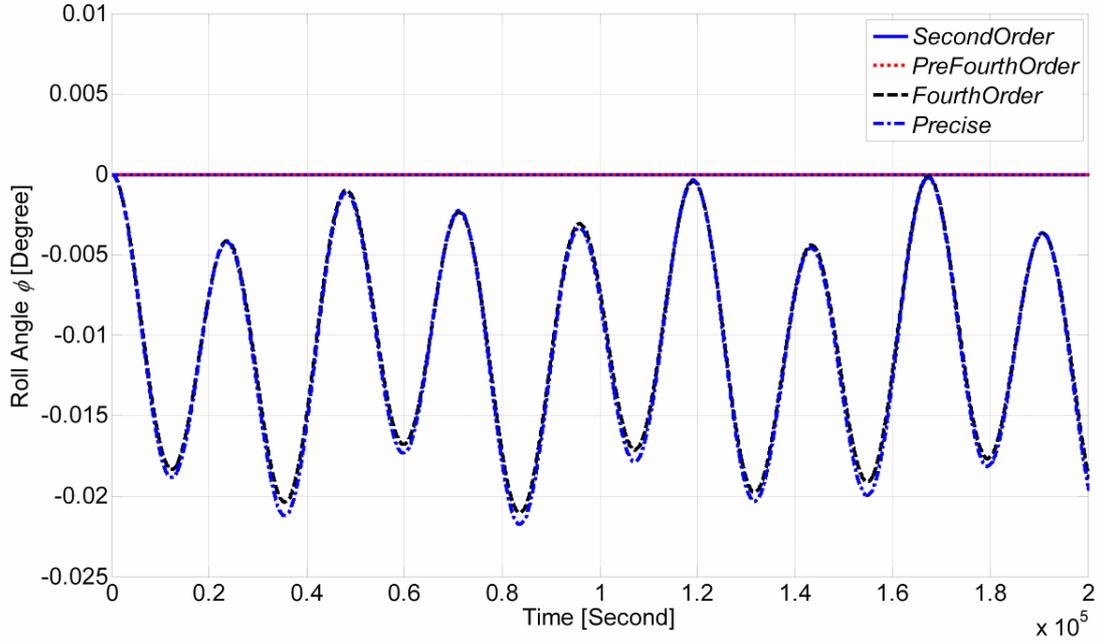

In the case of the previous fourth-order model *PreFourthOrder*, the Euler angles $\psi$, $\theta$ and $\phi$ are staying at zero, i.e. the spacecraft is at an equilibrium attitude, just as the second-order model *SecondOrder*. Namely, the previous fourth-order model has the same equilibria as the second-order model, and the non-central component of the gravity field of the asteroid has no effect on the location of the equilibrium in the previous fourth-order model.

However, the exact motion, as shown by Precise in Figures 5, 6 and 7, is small amplitude oscillation in all three axes, which is quite different from the previous fourth-order model *PreFourthOrder*. The amplitudes of the yaw and roll motions are the order of $10^{-2}$ degree, while the amplitude of the pitch motion is the order of $10^{-1}$ degree that would be a matter in the high-precision attitude dynamics. Therefore, the previous fourth-order model can not model the attitude motion very well.

From Figures 5, 6 and 7, we can see that our full fourth-order model *FourthOrder* fits the exact motion very well with the maximum error order of $10^{-3}$ degree in yaw and roll motions, and $10^{-2}$ degree in the pitch motion. Moreover, our full fourth-order model has the similar equilibria to the exact motion. Therefore, our full fourth-order model is more precise than the previous fourth-order model, and our model is precise enough for high-precision applications in attitude dynamics and control.

## Conclusion

In this paper, a full fourth-order gravity gradient torque model of spacecraft around a non-spherical asteroid is established. In this model, the gravity field of the asteroid is assumed to be 2nd degree and order with harmonic coefficients $C_{20}$ and

$C_{22}$. The inertia integrals of the spacecraft up to the fourth order are considered, that is an improvement with respect to previous fourth-order model. In previous fourth-order model, inertia integrals only up to the second order were considered. Through Taylor expansion, the mutual gravitational potential up to the fourth order is derived. Then the explicit formulations of the gravity gradient torque are obtained through the gravitational potential derivatives. We find that the third and fourth-order inertia integrals of the spacecraft appear in the third and fourth-order terms of the gravity gradient torque respectively along with the mass of the asteroid. In the previous fourth-order model, the third and fourth-order inertia integrals were not considered, thus the third-order terms and parts of fourth-order terms of the gravity gradient torque were neglected.

A numerical simulation is carried out to verify our full fourth-order model. In the numerical simulation, a special spacecraft consisted of 36 point masses, whose exact motion can be calculated, is considered. Simulation results show that the motion of the previous fourth-order model is quite different from the exact motion, while our full fourth-order model fits the exact motion very well. Our full fourth-order model has similar equilibria to the exact motion.

We can conclude that our full fourth-order gravity gradient torque model is more sound and precise than the previous fourth-order model, and our model is precise enough for high-precision attitude dynamics and control around asteroids.

## Appendix

The explicit formulations of the full fourth-order gravity gradient torque of the spacecraft $\tilde{T}_B$ are given as follows.

$$\tilde{T}_B^x = \frac{3\mu}{R^3}\bar{R}^y\bar{R}^z\left(I_{zz}-I_{yy}\right)+\frac{3\mu}{2R^4}\Big[\bar{R}^z\left(1-5\left(\bar{R}^y\right)^2\right)J_{yyy}-10\bar{R}^x\bar{R}^y\bar{R}^z J_{xyy}+10\bar{R}^x\bar{R}^y\bar{R}^z J_{xzz}+\bar{R}^y\left(5\left(\bar{R}^x\right)^2-1\right)J_{xxz}$$
$$+\bar{R}^z\left(1-5\left(\bar{R}^z\right)^2+10\left(\bar{R}^y\right)^2\right)J_{yzz}+\bar{R}^y\left(5\left(\bar{R}^z\right)^2-1\right)J_{zzz}-\bar{R}^y\left(1-5\left(\bar{R}^y\right)^2+10\left(\bar{R}^z\right)^2\right)J_{yyz}+10\bar{R}^x\left(\left(\bar{R}^y\right)^2-\left(\bar{R}^z\right)^2\right)J_{xyz}$$
$$+\bar{R}^z\left(1-5\left(\bar{R}^x\right)^2\right)J_{xxy}\Big]+\frac{5\mu}{2R^5}\Big[3\bar{R}^x\bar{R}^y\left(1-7\left(\bar{R}^z\right)^2\right)J_{xzzz}+\bar{R}^y\bar{R}^z\left(3-7\left(\bar{R}^z\right)^2\right)J_{zzzz}+3\bar{R}^y\bar{R}^z\left(1-7\left(\bar{R}^x\right)^2\right)J_{xxzz}$$
$$+21\bar{R}^y\bar{R}^z\left(\left(\bar{R}^z\right)^2-\left(\bar{R}^y\right)^2\right)J_{yyzz}+3\bar{R}^y\bar{R}^z\left(7\left(\bar{R}^x\right)^2-1\right)J_{xxyy}+\bar{R}^x\bar{R}^y\left(3-7\left(\bar{R}^x\right)^2\right)J_{xxxz}+\bar{R}^y\bar{R}^z\left(7\left(\bar{R}^y\right)^2-3\right)J_{yyyy}$$
$$+\left(3\left(\bar{R}^y\right)^2-3\left(\bar{R}^z\right)^2-7\left(\bar{R}^y\right)^4+21\left(\bar{R}^y\right)^2\left(\bar{R}^z\right)^2\right)J_{yyyz}+\left(3\left(\bar{R}^y\right)^2-3\left(\bar{R}^z\right)^2+7\left(\bar{R}^z\right)^4-21\left(\bar{R}^y\right)^2\left(\bar{R}^z\right)^2\right)J_{yzzz}$$
$$+\bar{R}^x\bar{R}^z\left(7\left(\bar{R}^x\right)^2-3\right)J_{xxxy}+3\left(\left(\bar{R}^y\right)^2-\left(\bar{R}^z\right)^2+7\left(\bar{R}^x\right)^2\left(\bar{R}^z\right)^2-7\left(\bar{R}^x\right)^2\left(\bar{R}^y\right)^2\right)J_{xxyz}+3\bar{R}^x\bar{R}^z\left(7\left(\bar{R}^y\right)^2-1\right)J_{xyyy}$$
$$-3\bar{R}^x\bar{R}^z\left(1-7\left(\bar{R}^z\right)^2+14\left(\bar{R}^y\right)^2\right)J_{xyzz}+3\bar{R}^x\bar{R}^y\left(1-7\left(\bar{R}^y\right)^2+14\left(\bar{R}^z\right)^2\right)J_{xyyz}\Big]$$
$$+\frac{5\mu}{2R^5}\left(I_{zz}-I_{yy}\right)\Big[3\tau_0\left(7\bar{R}^y\bar{R}^z\left(\boldsymbol{\gamma}\cdot\boldsymbol{R}\right)^2-2\bar{R}^y\gamma^z\left(\boldsymbol{\gamma}\cdot\boldsymbol{R}\right)-2\bar{R}^z\gamma^y\left(\boldsymbol{\gamma}\cdot\boldsymbol{R}\right)+\frac{2}{5}\gamma^y\gamma^z-\bar{R}^y\bar{R}^z\right)$$
$$+6\tau_2\left(7\bar{R}^y\bar{R}^z\left(\boldsymbol{\alpha}\cdot\boldsymbol{R}\right)^2-2\bar{R}^y\alpha^z\left(\boldsymbol{\alpha}\cdot\boldsymbol{R}\right)-2\bar{R}^z\alpha^y\left(\boldsymbol{\alpha}\cdot\boldsymbol{R}\right)+\frac{2}{5}\alpha^y\alpha^z\right)$$
$$-6\tau_2\left(7\bar{R}^y\bar{R}^z\left(\boldsymbol{\beta}\cdot\boldsymbol{R}\right)^2-2\bar{R}^y\beta^z\left(\boldsymbol{\beta}\cdot\boldsymbol{R}\right)-2\bar{R}^z\beta^y\left(\boldsymbol{\beta}\cdot\boldsymbol{R}\right)+\frac{2}{5}\beta^y\beta^z\right)\Big],\tag{A.1}$$

$$\tilde{T}_B^y = \frac{3\mu}{R^3}\bar{R}^x\bar{R}^z\left(I_{xx}-I_{zz}\right)+\frac{3\mu}{2R^4}\Big[\bar{R}^z\left(5\left(\bar{R}^x\right)^2-1\right)J_{xxx}+10\bar{R}^x\bar{R}^y\bar{R}^z J_{xxy}-10\bar{R}^x\bar{R}^y\bar{R}^z J_{yzz}+\bar{R}^z\left(5\left(\bar{R}^y\right)^2-1\right)J_{xyy}$$
$$+\bar{R}^x\left(1-5\left(\bar{R}^x\right)^2+10\left(\bar{R}^z\right)^2\right)J_{xxz}+\bar{R}^x\left(1-5\left(\bar{R}^z\right)^2\right)J_{zzz}-\bar{R}^z\left(1-5\left(\bar{R}^z\right)^2+10\left(\bar{R}^x\right)^2\right)J_{xzz}+10\bar{R}^y\left(\left(\bar{R}^z\right)^2-\left(\bar{R}^x\right)^2\right)J_{xyz}$$
$$+\bar{R}^x\left(1-5\left(\bar{R}^y\right)^2\right)J_{yyz}\Big]+\frac{5\mu}{2R^5}\Big[3\bar{R}^y\bar{R}^z\left(1-7\left(\bar{R}^x\right)^2\right)J_{xxxy}+\bar{R}^x\bar{R}^z\left(3-7\left(\bar{R}^x\right)^2\right)J_{xxxx}+3\bar{R}^x\bar{R}^z\left(1-7\left(\bar{R}^y\right)^2\right)J_{xyy}$$
$$+21\bar{R}^x\bar{R}^z\left(\left(\bar{R}^x\right)^2-\left(\bar{R}^z\right)^2\right)J_{xxzz}+3\bar{R}^x\bar{R}^z\left(7\left(\bar{R}^y\right)^2-1\right)J_{yyzz}+\bar{R}^y\bar{R}^z\left(3-7\left(\bar{R}^y\right)^2\right)J_{xyyy}+\bar{R}^x\bar{R}^y\left(7\left(\bar{R}^y\right)^2-3\right)J_{yyyz}$$
$$+\left(3\left(\bar{R}^z\right)^2-3\left(\bar{R}^x\right)^2-7\left(\bar{R}^z\right)^4+21\left(\bar{R}^x\right)^2\left(\bar{R}^z\right)^2\right)J_{xzzz}+\left(3\left(\bar{R}^z\right)^2-3\left(\bar{R}^x\right)^2+7\left(\bar{R}^x\right)^4-21\left(\bar{R}^x\right)^2\left(\bar{R}^z\right)^2\right)J_{xxxz}$$
$$+\bar{R}^x\bar{R}^z\left(7\left(\bar{R}^z\right)^2-3\right)J_{zzzz}+3\left(\left(\bar{R}^z\right)^2-\left(\bar{R}^x\right)^2+7\left(\bar{R}^x\right)^2\left(\bar{R}^y\right)^2-7\left(\bar{R}^y\right)^2\left(\bar{R}^z\right)^2\right)J_{xyyz}+3\bar{R}^x\bar{R}^y\left(7\left(\bar{R}^z\right)^2-1\right)J_{yzzz}$$
$$-3\bar{R}^x\bar{R}^y\left(1-7\left(\bar{R}^x\right)^2+14\left(\bar{R}^z\right)^2\right)J_{xxyz}+3\bar{R}^y\bar{R}^z\left(1-7\left(\bar{R}^z\right)^2+14\left(\bar{R}^x\right)^2\right)J_{xyzz}\Big]$$
$$+\frac{5\mu}{2R^5}\left(I_{xx}-I_{zz}\right)\Big[3\tau_0\left(7\bar{R}^x\bar{R}^z\left(\boldsymbol{\gamma}\cdot\boldsymbol{R}\right)^2-2\bar{R}^x\gamma^z\left(\boldsymbol{\gamma}\cdot\boldsymbol{R}\right)-2\bar{R}^z\gamma^x\left(\boldsymbol{\gamma}\cdot\boldsymbol{R}\right)+\frac{2}{5}\gamma^x\gamma^z-\bar{R}^x\bar{R}^z\right)$$
$$+6\tau_2\left(7\bar{R}^x\bar{R}^z\left(\boldsymbol{\alpha}\cdot\boldsymbol{R}\right)^2-2\bar{R}^x\alpha^z\left(\boldsymbol{\alpha}\cdot\boldsymbol{R}\right)-2\bar{R}^z\alpha^x\left(\boldsymbol{\alpha}\cdot\boldsymbol{R}\right)+\frac{2}{5}\alpha^x\alpha^z\right)$$
$$-6\tau_2\left(7\bar{R}^x\bar{R}^z\left(\boldsymbol{\beta}\cdot\boldsymbol{R}\right)^2-2\bar{R}^x\beta^z\left(\boldsymbol{\beta}\cdot\boldsymbol{R}\right)-2\bar{R}^z\beta^x\left(\boldsymbol{\beta}\cdot\boldsymbol{R}\right)+\frac{2}{5}\beta^x\beta^z\right)\Big],\tag{A.2}$$

$$\begin{aligned}
\tilde{T}_B^z =\ & \frac{3\mu}{R^3}\bar{R}^x\bar{R}^y\left(I_{yy}-I_{xx}\right)+\frac{3\mu}{2R^4}\bigg[\bar{R}^y\left(1-5\left(\bar{R}^x\right)^2\right)J_{xxx}+10\bar{R}^x\bar{R}^y\bar{R}^z J_{yyz}-10\bar{R}^x\bar{R}^y\bar{R}^z J_{xxz}+\bar{R}^x\left(5\left(\bar{R}^z\right)^2-1\right)J_{yzz} \\
& +\bar{R}^y\left(1-5\left(\bar{R}^y\right)^2+10\left(\bar{R}^x\right)^2\right)J_{xyy}+\bar{R}^x\left(5\left(\bar{R}^y\right)^2-1\right)J_{yyy}-\bar{R}^x\left(1-5\left(\bar{R}^x\right)^2+10\left(\bar{R}^y\right)^2\right)J_{xxy}+10\bar{R}^z\left(\left(\bar{R}^x\right)^2-\left(\bar{R}^y\right)^2\right)J_{xyz} \\
& +\bar{R}^y\left(1-5\left(\bar{R}^z\right)^2\right)J_{xzz}\bigg]+\frac{5\mu}{2R^5}\bigg[3\bar{R}^x\bar{R}^z\left(1-7\left(\bar{R}^y\right)^2\right)J_{yyz}+\bar{R}^x\bar{R}^y\left(3-7\left(\bar{R}^y\right)^2\right)J_{yyyy}+3\bar{R}^x\bar{R}^y\left(1-7\left(\bar{R}^z\right)^2\right)J_{yyzz} \\
& +21\bar{R}^x\bar{R}^y\left(\left(\bar{R}^y\right)^2-\left(\bar{R}^x\right)^2\right)J_{xxyy}+3\bar{R}^x\bar{R}^y\left(7\left(\bar{R}^z\right)^2-1\right)J_{xxzz}+\bar{R}^y\bar{R}^z\left(7\left(\bar{R}^z\right)^2-3\right)J_{xzzz}+3\bar{R}^y\bar{R}^z\left(7\left(\bar{R}^x\right)^2-1\right)J_{xxxz} \\
& +\left(3\left(\bar{R}^x\right)^2-3\left(\bar{R}^y\right)^2-7\left(\bar{R}^x\right)^4+21\left(\bar{R}^x\right)^2\left(\bar{R}^y\right)^2\right)J_{xxxy}+\left(3\left(\bar{R}^x\right)^2-3\left(\bar{R}^y\right)^2+7\left(\bar{R}^y\right)^4-21\left(\bar{R}^x\right)^2\left(\bar{R}^y\right)^2\right)J_{xyyy} \\
& +\bar{R}^x\bar{R}^y\left(7\left(\bar{R}^x\right)^2-3\right)J_{xxxx}+3\left(\left(\bar{R}^x\right)^2-\left(\bar{R}^y\right)^2+7\left(\bar{R}^y\right)^2\left(\bar{R}^z\right)^2-7\left(\bar{R}^x\right)^2\left(\bar{R}^z\right)^2\right)J_{xyzz}+\bar{R}^x\bar{R}^z\left(3-7\left(\bar{R}^z\right)^2\right)J_{yzzz} \\
& -3\bar{R}^y\bar{R}^z\left(1-7\left(\bar{R}^y\right)^2+14\left(\bar{R}^x\right)^2\right)J_{xyyz}+3\bar{R}^x\bar{R}^z\left(1-7\left(\bar{R}^x\right)^2+14\left(\bar{R}^y\right)^2\right)J_{xxyz}\bigg] \\
& +\frac{5\mu}{2R^5}\left(I_{yy}-I_{xx}\right)\bigg[3\tau_0\left(7\bar{R}^x\bar{R}^y\left(\boldsymbol{\gamma}\cdot\bar{\boldsymbol{R}}\right)^2-2\bar{R}^y\gamma^x\left(\boldsymbol{\gamma}\cdot\bar{\boldsymbol{R}}\right)-2\bar{R}^x\gamma^y\left(\boldsymbol{\gamma}\cdot\bar{\boldsymbol{R}}\right)+\frac{2}{5}\gamma^x\gamma^y-\bar{R}^x\bar{R}^y\right) \\
& +6\tau_2\left(7\bar{R}^x\bar{R}^y\left(\boldsymbol{\alpha}\cdot\bar{\boldsymbol{R}}\right)^2-2\bar{R}^y\alpha^x\left(\boldsymbol{\alpha}\cdot\bar{\boldsymbol{R}}\right)-2\bar{R}^x\alpha^y\left(\boldsymbol{\alpha}\cdot\bar{\boldsymbol{R}}\right)+\frac{2}{5}\alpha^x\alpha^y\right) \\
& -6\tau_2\left(7\bar{R}^x\bar{R}^y\left(\boldsymbol{\beta}\cdot\bar{\boldsymbol{R}}\right)^2-2\bar{R}^y\beta^x\left(\boldsymbol{\beta}\cdot\bar{\boldsymbol{R}}\right)-2\bar{R}^x\beta^y\left(\boldsymbol{\beta}\cdot\bar{\boldsymbol{R}}\right)+\frac{2}{5}\beta^x\beta^y\right)\bigg],
\end{aligned} \qquad (A.3)$$